\documentstyle[12pt,fleqn]{article}
\input epsf
\def\be{\begin{equation}}
\def\ee{\end{equation}}
\def\bea{\begin{eqnarray}}
\def\eea{\end{eqnarray}}

\def\fun#1#2{\lower3.6pt\vbox{\baselineskip0pt\lineskip.9pt
        \ialign{$\mathsurround=0pt#1\hfill##\hfil$\crcr#2\crcr\sim\crcr}}}
\def\re#1{{[\ref{#1}]}}
\def\reii#1#2{{[\ref{#1},\ref{#2}]}}
\def\reiii#1#2#3{{[\ref{#1},\ref{#2},\ref{#3}]}}

\def\thesection{}

\begin{document}
\thispagestyle{empty}
\renewcommand{\thefootnote}{\fnsymbol{footnote}}
\baselineskip=24pt
\begin{center}
{\Large \bf A BARYONIC CORRECTION TO GENERAL RELATIVITY}\\
\vspace{1.0cm}
\baselineskip=14pt

David E. Rosenberg\footnote{Electronic Mail: {\tt drosenbergmd@mcimail.com}}\\
{\em Section of astrophysics \\ 
Merkaz Hakolim, New Square, New York 10977}\\
\end{center}
\baselineskip=24pt

\begin{quote}
\hspace*{2em}The baryon overdensity and the matching of the big
bang explosion energy with gravitation can be solved by a
cyclical baryonic bounce model with correction to the
stress-energy tensor. Subtracting accretion energy from the CMBR
allows enough baryons in nucleosynthesis to close the
universe. Collapse to infinite density states must be prevented
by energy losses at  supranuclear densities. As long as the
Einstein tensor is coupled to the stress-energy tensor, any
quantum correction must involve an energy sink.
\vspace*{12pt}

\end{quote}


\newpage
\baselineskip=24pt
\setcounter{page}{2}
\renewcommand{\thefootnote}{\arabic{footnote}}
\addtocounter{footnote}{-3}

\thesection{\centerline{\bf I. INTRODUCTION: \\ 
Limits on general relativity}}
\setcounter{section}{1}
\setcounter{equation}{0}
\vspace{18pt}
General relativity was discovered early this century and  
twenty four years after its introduction, it was found to predict 
black holes. \re{Oppenheimer39a} Relativity has been
extrapolated to where stars, galaxies and the whole universe
could be compressed into a space smaller than an atom. There is
not one shred of evidence that the universe started at Planck
densities $\rho=10^{93}g/cm^3$ and temperatures
$T=10^{31}{ }^oK$. No high energy phenomena have been found from 
the first instant of creation. The nucleosynthesis of light
atomic nuclei ${}^4He$, ${}^2H$, and ${}^7Li$ took place around
densities of $10^5g/cm^3$ and temperatures of $\sim
10^{10}{}^oK$, according to accepted models \reii{Peebles66}{Wagoner67}. 
These conditions are the most extreme that has been confirmed in
the big bang. Thus general relativity, as applied to the
universe \re{Robertson36}, has been extrapolated eighty orders
of magnitude in density from points at which it has been
validated. Only for a homogeneous, isotropic universe, the field
equation has been simplified to the Friedmann equation      
\begin{equation}
{H^2}+ \bigl (K/a^2\bigr )\equiv \biggl ({\dot a\over
a}\biggr )^2 + (K/a^2)= {(8\pi\rho G)/3}\,,
\end{equation}    
where $H\equiv \dot a/a$ is the Hubble constant, which is
time dependent. G is the gravitational constant, $\rho$  is the
mass-energy density, p is the pressure and a(t) is the scale
factor of the universe, $\sim 10^{28}cm$, presently. There is
always a perfect fluid in the stress-energy tensor 
\begin{equation}
\bf {{T}}_{\alpha\beta}=\rho u_\alpha u_\beta+p(g_{\alpha\beta} +
u_\alpha u_\beta) \,,
\end{equation}
which ignores viscosity, shearing forces and subatomic effects
including differences between individual baryons and bulk
baryons, ${n_B}\gg 10^3$. 
Since it has been validated up to nuclear densities in pulsars, 
changes in the stress-energy tensor $\bf {{T}}_{\alpha\beta}$ 
at higher densities will be investigated.

\vspace{48pt}
\thesection{\centerline{\bf II. Theoretical changes necessary}}
\setcounter{section}{2}
\setcounter{equation}{0}
\vspace{18pt}
The Oppenheimer and Volkoff equations of
state \re{Oppenheimer39b} are used for neutron density matter
and neutron stars up to $3 x 10^{14} gm/cm^3$, 
\begin{eqnarray}
dm\over dr &= {4\pi r^2\rho} \cr\cr 
dp\over dr &= -{(\rho+p)(m+(4\pi r^3p))}/
{r(r-m)},
\end{eqnarray}  
where m is the mass within a given radius r. Since these
equations result from the field equation, 
information about the density change with pressure 
is also necessary. Neutron stars theoretically
have masses up to $5M_\odot$. 
Single neutrons have a compression energy about 300 MeV \re{Glendenning88}. 
Nuclear colliders start producing quark-gluon
plasma at energies over $2x10^{12}{ }^o K\approx 184MeV$. After
all the space in the neutron is eliminated 
$\rho >10^{17}gm/cm^3$, the net quantum effect of further
collapse and core compression must be a reversible energy
sink. Since nuclear pressures can't halt a gravitational
collapse, sufficient energy loss at supranuclear densities must
result in a stable configuration prior to quark formation. An
inhomogeneous collapse must stop when the compression energy
losses of core neutrons at peak $\rho\sim 10^{18}-10^{19} g/cm^3$
exactly match the gravitational energy, as shown in figure 1.

\vspace{48pt}
\begin{center}
\thesection{\bf III. {Resulting changes in our understanding}}
\end{center}
\setcounter{section}{3}
\setcounter{equation}{0}
\vspace{18pt}
Prior to the big bang, core densities slowly increased and the
energy sink of compressive losses rapidly overtook the collapse
energy by an overall mass-density $\rho\sim 10^{16}gm/cm^3$. 
If all the matter in the universe was in a spherical mass to
start, its radius was $\sim 10^{13}cm.$ As the density rose in
the core, the field disappeared and the pressure $p=\rho/3
\to 0$ in the stress-energy tensor as well.  By including this
energy loss, energy-momentum is conserved.  
With $\bf {{T}}$ and the vacuum energy $\lambda\approx 0$, 
an open universe existed during $t \le 0$. 
No singularities ever existed since there were no infinities in
energy, density or time. Accretion and other photons from
previous universes were very red shifted
by release into volumes $\gg$ today, so that they played no role
during the open universe period. Neutron compression energy
supplied $\approx 160MeV\approx 1.85x10^{12}{ }^oK$ which
propelled the farthest galaxies $\sim 0.5c$. After the bounce,
the metric changed to flat. 
There was no difference whether the early universe was closed 
or open \re{Misner73}. The extrinsic curvature $(6{\dot a}^2)/ a^2$ 
was much more important than the intrinsic
curvature $\pm 6/ a^2$ within any hyperspace of homogeneity,
since $\dot a^2$ was very large initially. The zones of influence
were too small to respond differently to negative or positive
spacetime curvatures.

The standard hot universe problems \re{Linde84}, can be
summarized and solved with the above correction.
The singularity problem follows from the scale factor of the
universe a(t) vanishes as $t\to 0$ and the energy density becomes
infinitely large. The inhomogeneity of matter with the energy
sink and red shifting of radiation prior to the big bang caused
the total energy-density $\to 0$.  

The flatness problem can be stated in several ways. The ratio
of the universe's mean mass density to the cold Einstein-de
Sitter universe 
\begin{equation}
{\rho/\rho_c}= {3H^2}/{8\pi\rho G} \,. 
\end{equation}
The Friedmann-Robertson-Walker(FRW) equation implies that this
ratio, which was proportional to curvature, was 
$1\pm 10^{-60}$ at the Planck era. The 'kinetic energy' $(\dot a/a)^2$ 
was equal to the gravitational mass-energy $8\pi\rho G/3$ ,
so that $k\approx 0$ in equation (1.1). 
Only a bounce mechanism by which the gravitational mass-energy 
was converted into kinetic energy could allow the universe to be
so flat. The unchanged nuclear state of the core allows 
this to happen without producing quarks. $\sim 160$ MeV was
sufficient to break the shell into billions of cold baryonic
masses $\le 10^{16}M_\odot$ . For mass M the gravitational radius
$R_g= GM/c^2$ then 
\begin{equation}
\rho= c^6/{G^3M^2} \,, 
\end{equation}
at black hole formation.  Thus primordial  holes could only be
formed from the expanding shell neutrons in masses $\ge 4M_\odot$
if $\rho_{max}\approx 3x10^{16}g/cm^3$. 
If this density can not be exceeded, then smaller black holes $<1M_\odot$
could not be formed, which would explain the missing Hawking
radiation \re{Halzen91}. 

The horizon problem has to do with areas in the initial
instant of creation that are too far from each other to have been
influenced by initial disturbances. A light pulse beginning at
t=0 will have travelled by time t, a physical distance 
\begin{equation}
l(t)= a(t){\int_0^tdt'}{ }a^{-1}(t')= 2t \,, 
\end{equation}
and this gives the physical horizon distance 
or Hubble radius dH. In a matter dominated universe without
vacuum energy $\lambda =0$, 
\begin{equation}
dH\approx 2H_0^{-1}\Omega_0^{-1/2}(1+z)^{-3/2} \,, 
\end{equation}
where $\Omega_0= {\rho/ \rho_c}$ in the present universe. 
This distance is compared with the radius L(t) of the region 
at time t which evolves into our currently 
observed area of the universe $\approx 10^{10}years$. 
Using a quark model near Planck conditions, this ratio $l^3/L^3$ 
is going to be very small, about $10^{-83}$. 
Since the average baryonic density initially is 
$\sim 10^{16}g/cm^3$  rather than Planck densities of
$10^{93}g/cm^3$, the horizon problem is diminished 
by a factor of $\sim 10^{77}$. Either the 
continuing loss of shell mass during $\bf{T}_{\alpha\beta}\approx
0$ or a major disturbance near equilibrium, will allow a nearly
simultaneous release of the stored neutron compression energy. 
Since state data on bulk nucleons is 
lacking, a reduction equation for a static system is extrapolated 
for compression losses of $E_{sink}=exp(\rho /2x10^{14})$ in the
energy term $\bf {T}_{\hat o\hat o}$. 

The homogeneity and isotropy problems arise due to the
postulated start of the universe in such a state. The
distribution of galaxies and clusters are not random on large
scales. A compilation of 869 clusters has shown a quasi-regular pattern
with high density regions separated by voids at intervals
$\approx 120 Mpc$. \re{Einasto97}. The CMBR has dipole
anisotropy not due to our Local Group motion \re{Lauer94}. 
The universe is not isotropic on its largest scales. It has long
been assumed that galaxy formation, which started after the
decoupling of matter and energy, grew by gravitational
amplification of small density fluctuations. 
With the Hubble space telescope, there is evidence that 
galaxies were assembled $z>4$ \re{Mo96}. Primordial galaxies,
composed of hot ${}^1H-{}^4He$ clouds orbiting the black hole
remnants of the cold shell, were already present
prior to decoupling of matter and energy $z\approx 1100$. 
As the universe expanded and the shell remnants 
separated, hydrogen was efficiently removed from intergalactic 
space down to the Gunn-Peterson ${}^1H$ limit, and attenuated the
CMBR temperature gradients as follows.  Hot electrons upscattered
the redshifted photons emitted by orbiting hydrogen deeper in the
protogalactic wells. With decoupling, there are three types
of scattering which accomplished this. \re{Padmanabhan93}
Thompson scattering by itself can not help thermalization because
there is no energy exchange between the photons and electrons. If 
\begin{equation}
\sigma_T= 8\pi /3(e^2/m)^2 \,, 
\end{equation}
is the Thompson scattering cross section, 
then the mean-free-path for a photon between collisions is 
\begin{equation}
\lambda_{\gamma}= (\sigma_T n_e)^{-1} \,, 
\end{equation}
where $n_e$ is the number density of electrons. 
While traveling a distance l, the photon will perform a 
random walk and undergo N collisions where 
$N^{1/2}\lambda_{\gamma}=l$.  Since Compton scattering will not
change the number of photons, it will not create a Planck
spectrum. Free-free absorption at a frequency
$\omega$, is given by \re{Padmanabhan93}. 
\begin{equation}
t_{ff}\cong {3(6\pi mT)^{1/2}m\omega^3}/
{(32e^6n_e^2\pi^3)/(1-e^{-\omega/T})} \,. 
\end{equation}
For photons with a frequency $\omega\approx T$ in electron volts, 
\begin{equation}
t_{ff}= 2x10^{14}sec(\Omega_Bh^2x_e)^{-2}T^{-5/2} \,.
\end{equation}
For ionization fraction $x_e\approx 1$, 
\begin{equation}
t_{ff}/H^{-1}\approx (T/1.9x10^4eV)^{-1/2}(\Omega_Bh^2)^{-2} \,. 
\end{equation}
Thompson scattering increases the effective path length 
for photon absorption of free-free scattering 
\begin{equation}
\bar t= 1.1x10^{11}sec.{ }T^{-11/4}(\Omega_Bh^2X_e)^{-3/2} \,. 
\end{equation}
With primordial galaxies, free-free can dominate over Compton scattering 
between 90eV-1eV, lead to true thermalization and diminish
temperature gradients in the CMBR. In FRW geometry, radiation
energy $\rho_R\propto a^{-4}$ and $T\propto a^{-1}$. An increase
in a(t) from $10^{13}cm.$ to $10^{28}cm.$ today would cause the
corresponding temperature of CMBR would be $.00185{}^oK$, 
without the accretion energy released from the previous universe.
Big bang photons are thus the small tail of the thermal spectrum
near absolute zero. See for example \re{Zeldovich83} which
discusses reasons for this tail. The smooth Planck spectrum at
$2.73{}^oK$ with $\delta T/T\sim 10^{-5}$ was released by
accretion during the collapse of the previous universe at
$a(t)\sim 10^{22}cm.$, as shown in figure 2. The photon number
density $cm^{-3}$ 
\begin{equation}
n_\gamma =2.038x10^{28}T_9^3 \,, 
\end{equation}
where $T_9$ is the temperature in units of $10^9{ }^oK$. 
Therefore the photon density of the big bang is $1.29x10^{-7}$.
rather than 422. This changes the baryon/photon ratio to
\begin{equation}
\eta= 87.6{\Omega_B}h^2 \,, 
\end{equation}
where h is the Hubble constant in units of $100km.sec^{-1}Mpc^{-1}$. 
The explosion mechanism and $\eta$ are similar to that of a supernova.
The hot baryon to photon ratio must be multiplied by the cold
baryon factor CBF plus one to obtain the total baryon/photon ratio 
\begin{equation}
\eta_{hot}(CBF+1)=\eta_{total} \,. 
\end{equation}
A total $\eta\ge 100$ will definitely close the universe with
baryons. With the kind assistance of Edward W.(Rocky)Kolb, 
the nucleosynthesis program NUC123 of Larry Kawano was modified. 
Cold baryons were calculated by multiplying the hot baryon
density thm(9) in subroutine therm by the cold baryon multiplier. 
This was added to the total energy density thm(10) and thus to 
the Hubble constant. The program was compiled using the fortran77
compiler of the Absoft Corporation with the Vax compatibility
option. A double precision option for all floating point
variables and disabling of overflow checking allowed calculations
with hot $\eta >1$. 
Using cold baryons, neutrino degeneration and $\eta$ as
variables,it was found that $\eta=10^{-7}$, 
a cold baryon multiplier $10^9$ and an 
electron neutrino chemical potential $\xi_{\nu_e} =1.8$ 
gave a D or ${}^2H/H=1.9x10^{-4}$ and a ${}^4He/H=.246$. 
Using cold baryons allowed yields of ${}^2H/H>10^{-4}$. 
The deuterium fraction increased with increasing
cold baryons. The ${}^4He$ yields decreased with increasing
electron neutrino chemical potential by reducing the neutron 
to proton ratio at freeze out, as first noted \re{Wagoner67}.
Doubling the cold baryons gave a ${}^2H/H=2.07x10^{-4}$ without
change to other yields. The other yields were 
\begin{eqnarray}
{}^3H&=&5.35x10^{-7}\quad{}^3He=1.4x10^{-5}\quad{}^7Li=1.4x10^{-10} \cr
N&=&6.8x10^{-8}\quad{}^6Li=4.2x10^{-14}\quad{}^7Be=4.1x10{-11} \cr
{}^8Li+up&=&1.7x10^{-15},. 
\end{eqnarray}
These are all compatable with the standard nucleosynthesis yields
except the nitrogen fraction which has $N=5.6x10^{-16}$. 
The low estimate deuterium fraction now in favor $(1-2)x10^{-5}$,  
could be made with the same neutrino degeneration, $\eta=10^{-6}$ 
and cold baryon factor of $10^9$ as well as $\eta =10^{-7}$ 
and a cold baryon factor of $2x10^8$. The last case probably
hasn't sufficient baryons to close the universe. 

Galaxy formation problems \re{Peebles90} are greatly
simplified. An explosive universe with galaxy formation will fit
the large scale galactic pattern \re{Weinberg89}. 
Although the Jeans mass is thought to be the point 
at which gravity overcomes pressure to form galaxies, massive
rotating primordial black holes may be necessary for galactic
structure. In the Tully-Fisher relation 
\begin{equation}
V_c= 220(L/L_\star)^{.22} \,,
\end{equation}
and Faber-Jackson 
\begin{equation}
V_c= 220 (L/L_\star)^{.25} \,, 
\end{equation}
where $V_c$ is the circular velocity $km/sec$ and $L_\star$ 
is the characteristic galaxy luminosity. The former relation is
for velocities in the dark halo of spiral galaxies and the latter
for star velocity dispersion in central parts of elliptical
galaxies \re{Peebles93}. Rotational energy is a function of
$MV_c^2$. Galactic brightness results from ${}^1H$ mass, $M_{galaxy}$. The
black hole capturing cross section 
\begin{equation}
\sigma_{capt.}=16\pi M^2/ \beta^2 \,, 
\end{equation}
where $\beta$ is the particle velocity relative to
light \re{Misner73}. Because of the ${}^1H$ capture by
primordial black holes, the brightness is proportional to the
central nuclear mass $M_{nucleus}^2$. With $M_{nucleus}^2V_c^4=$ constant, 
$M_{nucleus}V_c^2$ is constant related to the rotational 
energy imparted prior to the big bang. 
Thus Tully-Fisher can relate the stellar galactic
mass and luminosity to the depth of the dark matter potential
well and asymtotic circular speed. Due to the capture mechanism
of ${}^1H$, the black hole nuclear mass $M_{nucleus}\propto M_{galaxy}$.
Galaxy formation never involved collapse dynamics
with its different post collapse densities, circular speeds and
disk assymetries.   

The quantization of galactic redshifts found in even
multiples of 37km./sec. by W.G. Tifft \reiii{Tifft74}{Tifft76}{Tifft84}
and other workers \reiii{Broadhurst90}{Arp90}{Karlsson71} and also 
\reii{Depaquit85}{Guthrie91} is persuasive evidence that the cold 
baryonic shell, which formed galactic nuclei and quasars, was
present already at the big bang. Its different layers 
received different energies from the hot expanding core, 
even producing supermassive black holes.
Near Abell 3627 there is a mass  $5x10^{16}M_\odot$ , the Great
Attractor \re{Kraan-Korteweg96}, which must result from a large
initial homogeneity. It may be near the original site of the big
bang. The explosion mechanism described here is apparently that
in the Hebrew Bible.

The baryon asymmetry problem has been stated as to why there 
are many more baryons than antibaryons. Baryon-antibaryon pairs 
are only created from a vacuum at energies $>10^{13}{ }^oK$, which
is higher than the $160MeV\approx 1.85x10^{12}{ }^oK$ core
temperature. Extreme energy phenomena such as domain walls, monopoles,
gravitinos and symmetry breaking were not reached in big bang.    

\vspace{48pt}
\thesection{\centerline{\bf IV. A cyclical universe}}
\setcounter{section}{4}
\setcounter{equation}{0}
\vspace{18pt}

Although equation 1 is cyclical, it is valid only for a 
universe that is isotropic and homogeneous i.e. a perfect fluid.
In figure 2, the maximum scale factor $a_{max}$ of the universe
is equal to the gravitational radius 
\begin{equation}
R_g=GM/c^2\sim 10^{29}cm \,. 
\end{equation}
After $a_{max}$ was reached, the galaxies were blue shifted 
as they reconverged. When a(t) was $10^6$ smaller than today, 
the proportionately higher CMBR tore neutrons 
and protons from nuclei. In the center was a growing black hole 
resulting from merging galactic nuclei. Stars and galaxies were 
accreted onto this supermassive black hole in a massive thick
disk. Once the mass of this black hole exceeded the size of an
average galactic nucleus $\sim 10^8M_\odot$, 
tidal forces were no longer capable of tearing a star apart
before it entered $R_g$ with relatively little radiative 
losses \re{Frank92}. The collapsing scale factor a(t) forced 
all the matter and released energy inside the growing 
$R_g$ in a Schwarzschild gometry. Then 
$R_g\to 0$ as the spacetime propogation of the core energy losses
slowly reduced the potential barrier of the supermassive black
hole.    

\vspace{48pt}
\thesection{\centerline{\bf V. DISCUSSION}}
\setcounter{section}{5}
\setcounter{equation}{0}
\vspace{18pt}

Although classical general relativity has been confirmed to one
part in $10^{12}$, it must break down prior to the infinite
densities of singularities. There is no reason why a small mass
$>4M_\odot$ can contract to a singularity while the mass of
universe explodes into the big bang. If a star surface lies
entirely inside the $R_g$, classical relativity concludes from
Kruskal-Szekeres diagrams that it must collapse to a singularity
or faster than the speed of light. Here coordinate reversal
occurs, $\partial/\partial r$ is timelike $(g_{rr}<0)$ and proper
time at the surface 
\begin{equation}
\tau= -\int^R\bigl [g_{rr}\bigr ]^{1/2}dr + constant \,. 
\end{equation}
In order to allow a big bang, a reduction in the 
stress-energy tensor must occur before enormous densities and
energies are reached inside $R_g$. As
$\bf {{T}}\to 0$, the impetus for further collapse stops with
eventual elimination of the future event horizon. After
equilibrium is established, there is re-reversal of the time
coordinate and no further reduction in size. 
The quantum requirement that $\bf {{T}}>0$, 
will not be violated as it will approach zero on the positive
side. A solution to the covariant perturbation problem for quantum
gravity would be as follows. The spacetime metric $g_{ab}$ is divided
into a flat Minkowski component $\beta_{ab}$ and its deviation 
$\gamma_{ab}$, where $(M,{}^og_{ab})$ is a solution to the field
equation. The field equation can be seen as an equation 
for a self interacting spin-2 field $\gamma_{ab}$ in Minkowski spacetime.
In the first order $\gamma_{ab}$ is a free spin-2 equation with much
gauge arbitrariness which can be expanded into a perturbation series 
for non-abelian gauge fields. Although this part is non-renormalizable, 
the energy sink correction eliminates this term at high energies leaving
the background metric $\beta_{ab}$ which satisfies causality conditions.
The quantum mechanism by which the energy sink suppresses vibratory and
other modes remains to be elucidated.
The problem of evaporation for black holes under a solar
mass due to quantum particle creation with violation of lepton
and baryon conservation is avoided. Naked and all other
singularities are mathematically eliminated. Black holes can
eventually influence their surroundings to achieve thermal
equilibrium. Thus there is no loss of quantum coherence as the
final black hole state will be a pure one and the scattering
matrix S deterministic. Supernovas $<4M_\odot$, when collapsing to  
the same limiting density, will bounce without blackhole formation. 
A supranuclear equation of state based on
actual data (which does not yet exist) or more accurate
primordial deuterium ambundance would better determine the shell
to core mass ratios and the bounce temperature.

\vspace{36pt}
Acknowledgements. This paper would not have been possible without
many helpful comments from J. Schiff, E.Kolb, D. Lindley, D. Spergel 
and I. Klebanoff. This paper was copyrighted 1993,1997.



\begin{picture}(400,50)(0,0)

\put (50,0){\line(350,0){300}}
\end{picture}

\vspace{0.25in}

\def\labelenumi{[\theenumi]}
\frenchspacing
\def\prl{{{\em Phys. Rev. Lett.\ }}}
\def\prd{{{\em Phys. Rev. D\ }}}
\def\pl{{{\em Phys. Lett.\ }}}

\begin{enumerate}

\item\label{Arp90} Arp, H. Ast. Astrophys. 229, 93-98 (1990)

\item\label{Broadhurst90} Broadhurst, T.J., Ellis, R.S., Koo,
D.C.and Szalay, A.S. Nature 343, 726-728 (1990)

\item\label{Depaquit85} Depaquit, S., Pecker, J.C. and Vigier,
J.P. Astr. Nachr. 306, 7 (1985)

\item\label{Einasto97} Einasto, J. et al. Nature 385, 139-141
(1997) 

\item\label{Frank92} Frank, J., King, A. and Raine, D. Accretion
Power in Astrophysics (Cambridge University Press, Cambridge
England, 1992) 171-265

\item\label{Glendenning88} Glendenning, N.K. Phys. Rev. C 37,
2733-2742 (1988)

\item\label{Guthrie91} Guthrie, B.N.G. and Napier, W.M. Mon. Not.
R. Astr. Soc. 253, 533-544 (1991)

\item\label{Halzen91} Halzen, F., Mac Gibbon J.H. and Weeks, T.C.
Nature 353, 807-814 (1991)

\item\label{Karlsson71} Karlsson, K.G. Ast. Astrophys. 13, 333-335
(1971)

\item\label{Kraan-Korteweg96} Kraan-Korteweg, R.C. et al. Nature
379, 519-521 (1996)

\item\label{Lauer94} Lauer, T.R. and Postman, M. Astrophys. J. 425,
418-438 (1994)

\item\label{Linde84} Linde, A.D. Rep. Prog. Phys. 47, 925-986 (1984)

\item\label{Misner73} Misner, C.W., Thorne, K.S. and Wheeler, J.A.
Gravitation (W.H. Freeman and Co., New York, 1973) 655-679,
703-816, 872-915

\item\label{Mo96} Mo, H.J., Fukugita, M. astro-ph 9604034

\item\label{Oppenheimer39a} Oppenheimer, J.R. and Snyder, H. Phys.
Rev. 56, 455-459 (1939)

\item\label{Oppenheimer39b} Oppenheimer, J.R. and Volkoff, G.M.
Phys. Rev. 55, 374-381 (1939)

\item\label{Padmanabhan93} Padmanabhan, P. Structure Formation In
The Universe (Cambridge University Press, Cambridge, England,
1993)70-73, 108-112, 217-247, 325-352

\item\label{Peebles66} Peebles, P.J.E. Astrophys. J. 146, 542-552
(1966)

\item\label{Peebles90} Peebles, P.J.E. and Silk, J. Nature 346,
233-239 (1990)

\item\label{Peebles93}Peebles, P.J.E. Principles of Physical
Cosmology (Princeton University Press, Princeton, N.J.,1993)
45-54, 527-564

\item\label{Robertson36} Robertson, H.P. Astrophys. J. 83, 187-201,
257-271 (1936)

\item\label{Tifft74} Tifft, W.G. The Formation and Dynamics of 
Galaxies, IAU Symp. No. 58, ed. Shakeshaft, J.R. (Reidel,
Dordrecht, 1974) 243

\item\label{Tifft76} Tifft, W.G. Astrophys. J. 206, 38-56 (1976)

\item\label{Tifft84} Tifft, W.G. and Cocke, W.J. Astrophys. J. 287,
492-502 (1984)

\item\label{Wagoner67} Wagoner, R.V., Fowler, W.A. and Hoyle, F.
Astrophys. J. 148, 3-49 (1967)

\item\label{Weinberg89} Weinberg, D.H., Ostriker, J.P. and Dekel, A.
Astrophys. J. 336, 9-45 (1989)

\item\label{Zeldovich83} Zel'dovich, Y.A. and Novikov, I.D. The
Structure and Evolution of The Universe 
(University of Chicago Press, Chicago, Ill., 1983) 209-233

\end{enumerate}
\end{document}